\input phyzzx
\baselineskip=22pt
\input epsf
\def\scrim{{\cal I}^-}
\def\scrip{{\cal I}^+}

\def\of{{\overline f}}
\def\oh{{\overline h}}
\def\oa{{\overline a}}
\def\tR{{\tilde R}}
\def\ty{{\tilde y}}
\def\tU{{\tilde U}}
\def\tV{{\tilde V}}
\def\tu{{\tilde u}}

\def\w{{\omega}}
\rightline{UATP-99/04}
\vskip 0.1in
\centerline{\fourteenrm Quantum Radiation from Black Holes and Naked Singularities}
\centerline{\fourteenrm in Spherical Dust Collapse} 
\vskip 0.25in
\centerline{{\caps T. P. Singh}\footnote{\dagger}{Email: tpsingh@tifr.res.in}}
\centerline{\it Tata Institute of Fundamental Research}
\centerline{\it Homi Bhabha Road, Mumbai 400 005, India.}
\vskip 0.25in
\centerline{{\caps Cenalo Vaz}\footnote{\ddagger}{Email: cvaz@ualg.pt.}}
\centerline{\it UCEH, Universidade do Algarve}
\centerline{\it Campus de Gambelas}
\centerline{\it P-8000 Faro, Portugal.}
\vskip 0.25in

\centerline{\caps Abstract}
\vskip 0.25in

\noindent A sufficiently massive collapsing star will end its life as a spacetime 
singularity. The nature of the Hawking radiation emitted during collapse depends  
critically on whether the star's boundary conditions are such as would lead to the 
eventual formation of a black hole or, alternatively, to the formation of a naked 
singularity. This latter possibility is not excluded by the singularity theorems. 
We discuss the nature of the Hawking radiation emitted in each case.  We justify 
the use of Bogoliubov transforms in the presence of a Cauchy horizon and show 
that if spacetime is assumed to terminate at the Cauchy horizon, the resulting 
spectrum is thermal, but with a temperature different from the Hawking temperature.
\vskip 0.25in

\noindent{\caps PACS:} 04.20.Dw, 04.62+v, 04.70-s
\vfill \eject

\noindent{\bf 1. Introduction.}

There are by now many known examples of formation of naked singularities in
spherical gravitational collapse in classical general relativity.${}^{[1,2]}$ Whereas these
examples do not necessarily invalidate the Cosmic Censorship Hypothesis,${}^{[3]}$ 
it is interesting to ask what a star forming a naked singularity would look 
like to a distant observer. Since visible regions of very high curvature develop during 
the collapse, it can be expected that quantum effects will play a significant role in 
determining the evolution of the star. Furthermore, it should be possible to describe 
these quantum effects using techniques from quantum field theory up to the time when 
curvatures approach Planck scales.

A principal issue is a comparison between the Hawking evaporation of a star that forms a 
black hole, and the corresponding quantum evaporation of a star that forms a naked 
singularity. Studies of this problem can be divided into two classes; one in which the 
vacuum expectation value (VEV) of the energy momentum tensor of a quantized field is 
calculated in the background of the classical collapsing star using the trace 
anomaly, and the other in which the VEV of the radiation flux and spectrum of the radiation 
is calculated asymptotically in the geometric optics approximation, using Bogoliubov 
coefficients.${}^{[4]}$

A few studies have been carried out in recent years to calculate the VEV of the quantized 
stress tensor in spacetimes which evolve to naked singularities. The central idea here is 
to investigate the behavior of the VEV in the approach to the Cauchy horizon. It has been 
found in all examples of shell focussing naked singularities that the flux of radiation 
diverges as the Cauchy horizon is approached.  The divergence of the outgoing flux of the 
quantum field on the Cauchy horizon would suggest that the back-reaction ultimately prevents 
the naked singularity from forming. Furthermore, a divergent flux could in principle be 
measured by a distant observer if such objects were to occur in nature.

Perhaps the first investigation in this context was due to Ford and Parker,${}^{[5]}$ who 
calculated the outgoing flux of a quantized massless scalar field in the spacetime of a 
collapsing spherical dust cloud which develops a shell-crossing naked singularity. The flux 
was calculated using the geometric optics approximation, and remains finite in the approach 
to the formation of the naked singularity. Another early calculation was by Hiscock et 
al.,${}^{[6]}$ who computed the outgoing flux for a massless scalar field in the two dimensional
self-similar Vaidya spacetime (obtained by suppressing angular coordinates in the spherical 
spacetime) which evolves to a shell-focusing  naked singularity. In this instance, the flux 
diverges on the Cauchy horizon. In recent works${}^{[7]}$ we calculated the outgoing flux 
in the background spacetime of a spherical self-similar dust cloud with a naked singularity, 
using both the geometric optics approximation (analogous to Ford and Parker${}^{[5]}$) and the 
trace anomaly (analogous to Hiscock et al.${}^{[6]}$). Again, the flux diverges on the Cauchy 
horizon.

As mentioned above, apart from the calculation of the quantum stress tensor, a calculation of 
the spectrum of the created particles is also of great interest. Since the calculation of the 
stress tensor is local, it can be carried out using standard techniques of quantum theory. 
However, the presence of the Cauchy horizon raises subtle issues in the 
calculation of the spectrum, and the purpose of the present paper is to address some of these 
issues.

Consider the Penrose diagram for a collapsing star which develops a naked
singularity (figure 1).
\vskip 0.1in
\centerline{\epsfysize=1.5in \epsfbox{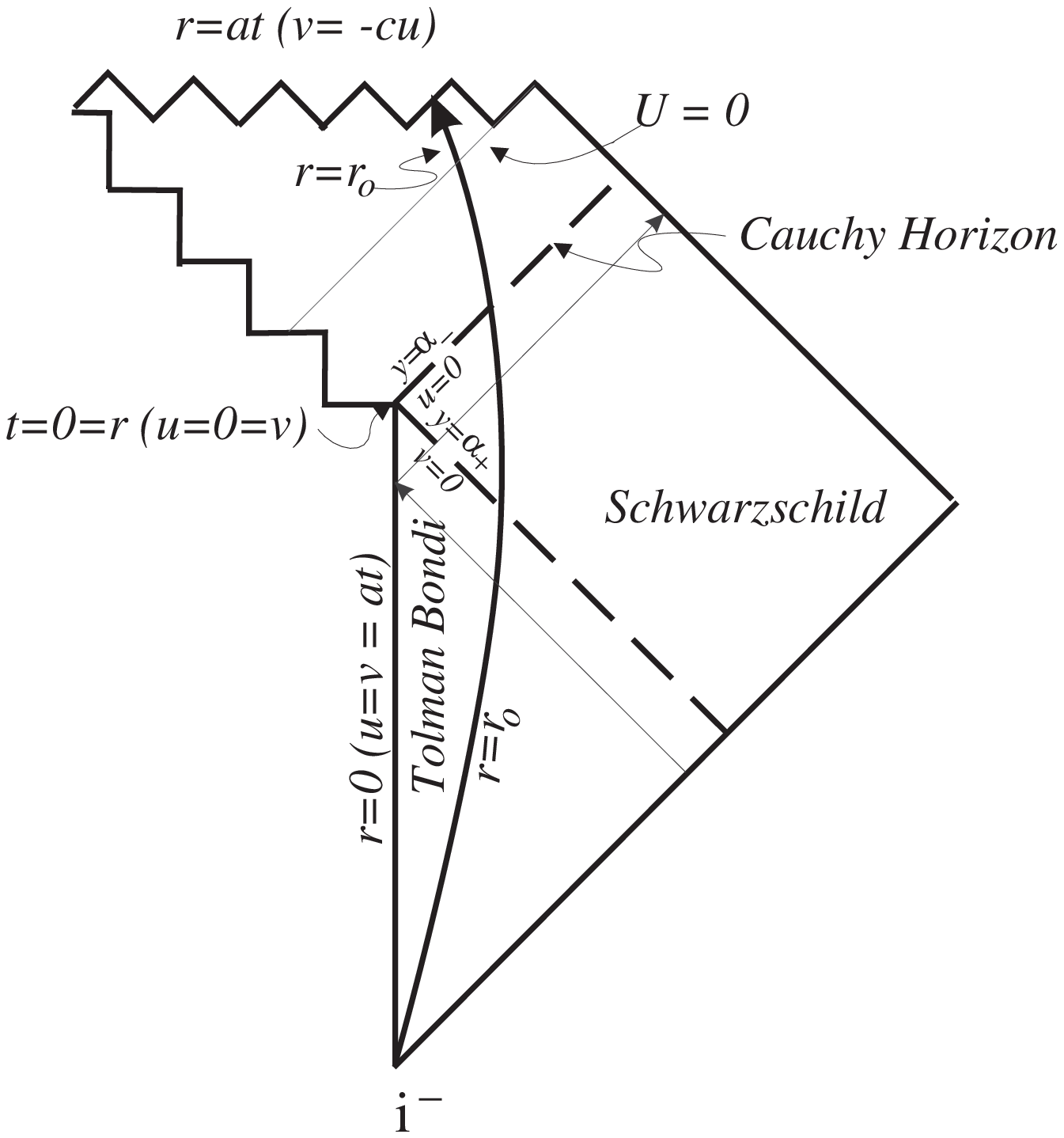}}
\centerline{\Tenpoint Figure 1. Formation of a naked singularity}
\vskip 0.1in

\noindent The presence of a Cauchy horizon implies that the region of $\scrip$ to the future of 
the Cauchy horizon is exposed to the naked singularity. Hence one might conclude that a complete 
set of modes cannot be defined on $\scrip$, and the standard Bogoliubov calculation cannot be 
carried out${}^{[6]}$.

There are two ways out of this apparent obstacle. One is to consider the very real possibility 
that a star collapsing to a naked singularity does not destroy the whole universe. $\scrip$ 
continues to be well-defined beyond the Cauchy horizon, and since no ingoing modes reach out to 
this part of $\scrip$, the outgoing modes in this region can be set to zero. It has been shown 
by us that the resulting spectrum is non-thermal.${}^{[8]}$ In section 2 we give a justification 
for this approach by showing that the total radiation computed from this spectrum indeed equals 
the integrated flux obtained from the stress tensor.

A second way out of the apparent obstacle is to assume that the Cauchy horizon is actually the 
end of spacetime, {\it i.e.,} that the analytical continuation which defines the space time to 
its future is not physical. We show in section 4 that in this case the spectrum is black-body, with a 
temperature different from that for the Hawking black hole. This case is analogous to that of a 
marginally naked singularity (treated in ref. [6]), {\it i.e.,} one in which the Cauchy horizon 
coincides with the event horizon.

Another issue in the calculation of the spectrum concerns the existence of a complete orthonormal 
basis set of infalling waves on the Cauchy horizon. In principle, the quantum field should be 
expressible in complete bases on $\scrip \cup {\cal H}_C$, where ${\cal H}_C$ is the Cauchy horizon. 
There is an obvious difficulty in constructing a basis of infalling waves on ${\cal H}_C$ on account 
of the central singularity, and so there is an essential ambiguity in evolving the field in the 
future of the Cauchy horizon. This ambiguity is inconsequential, for it would be relevant 
if one were interested in constructing outgoing wave packets on $\scrip$ in the future of the Cauchy 
horizon, from infalling waves on $\scrim$. No such outgoing packets exist, however, because all probes 
coming in from $\scrim$ at such late advanced times would be absorbed by the singularity. We are 
therefore only interested in complete orthonormal sets of solutions to the wave equation on hypersurfaces 
in the past of the Cauchy horizon. These are well defined. Moreover, the spectrum on $\scrip$ will 
be independent of any particular choice of infalling basis states.

We will compare the features of the Hawking radiation from black holes and naked singularities 
via a model of self-similar collapse based on inhomogeneous dust which is described in section 3. 
Our reason for using this particular model is twofold: firstly, the causal structure of the spacetime 
is well understood${}^{[9]}$ and secondly, the model exhibits both classical black hole and naked 
singularity end states, thus allowing for a comparison between the behaviors of each case within a 
unified picture. We do not expect any of the conclusions to be heavily dependent on the model itself, 
based as they are on general arguments that would apply whenever black holes or naked singularities 
are formed. 
\vskip 0.25in

\noindent{\bf 2. Relationship between the flux and spectrum.}

Consider the propagation of a scalar field in some spherically symmetric, asymptotically flat 
background spacetime. For convenience we consider a massless scalar field, though massive 
fields as well as fields of arbitrary spin may be treated using the same techniques. We follow 
closely the treatment of Ford and Parker${}^{[5]}$ (see also Birrell and Davies in ref. [4]). In this 
spacetime, we will assume that the radial null rays define a one-to-one mapping between 
a portion of past null infinity, $\scrim$, and some region of future null infinity, $\scrip$.
We will also assume that a complete basis set of null infalling rays may be defined on 
$\scrim$ and a complete basis of null outgoing rays defined on $\scrip$. Let $t,r,\theta$ and 
$\phi$ define a quasi-Minkowskian coordinate system in the asymptotic region and let $\tU = t-r$ 
and $\tV=t+r$ be the the null coordinates there. We imagine that a null incoming ray, at 
$\tV = $ constant, originates on $\scrim$ and propagates through the spacetime geometry, turning 
into a null outgoing ray, $\tU=$ constant, on $\scrip$ with value $\tU=F(\tV)$. In the time reversed 
situation, one could trace a null ray, $\tU=$ constant, on $\scrip$ into the past. Such a ray 
would have originated at $\tV=G(\tU)$ on $\scrim$, so that $G(\tU)$ is the inverse of $F(\tV)$.
In Minkowski space, for example, $F(\tV)=\tV$ and $G(\tU)=\tU$. If the  functions, $F(\tV)$, are known or 
can be determined, one considers positive energy solutions of the massless wave equation which 
have the following asymptotic form
$$\of_{\omega l m}~~ \sim ~~ {1 \over {\sqrt{4\pi\omega}r}}\left[e^{-i\omega \tU}~ +~ e^{-i\omega F(\tV)}
\right] Y_{lm}(\theta,\phi),\eqno(2.1)$$
which corresponds to an outgoing plane wave on $\scrip$ and is normalized on a spatial 
hypersurface in the asymptotically flat ``out'' region according to
$$\langle \of_{\omega l m}, \of_{\omega' l' m'}\rangle~~ =~~ \delta(\omega-\omega') \delta_{ll'} 
\delta_{mm'},\eqno(2.2)$$
with the inner product being defined by
$$\langle \of, \oh\rangle~~ =~~ -i \int_\Sigma d\Sigma^\mu {\sqrt{g_\Sigma}}~ \left[\of 
(\partial_\mu \oh^*)~ -~ (\partial_\mu \of) \oh^*\right],\eqno(2.3)$$
$\Sigma$ being the hypersurface. Wavepackets  formed from the $\of_{\omega l m}$ 
are outgoing plane waves at late times and incoming at early times in accordance with popagation by 
geometrical optics, and any positive energy solution of the scalar wave equation that is outgoing on $\scrip$ 
can be written as a wavepacket formed from the $\of_{\omega l m}$. An equivalent expression can be 
given for wavepackets that are incoming plane waves on $\scrim$. These will be formed from 
$$f_{\omega l m}~~ \sim~~ {1 \over {\sqrt{4\pi\omega}r}}\left[e^{-i\omega \tV}~ +~ e^{-i\omega G(\tU)}
\right] Y_{lm}(\theta,\phi).\eqno(2.4)$$
A quantum field, $\phi$, may therefore be expanded in either basis
$$\eqalign{\phi~~ &=~~ \sum_{lm} \int_0^\infty d\omega [\oa_{\omega l m} \of_{\omega l m}~ +~ 
\oa^\dagger_{\omega l m} \of^*_{\omega l m}]\cr &=~~ \sum_{lm} \int_0^\infty d\omega [a_{\omega l m} 
f_{\omega l m}~ +~ a^\dagger_{\omega l m} f^*_{\omega l m}],\cr}\eqno(2.5)$$
in terms of the annihilation and creation operators, $a_{\omega l m}$ and $\oa_{\omega l m}$, and 
their hermitean conjugates. The vacuum defined by the $a_{\omega l m}$, according 
to $a_{\omega l m}|0\rangle = 0$, is the ``in'' vacuum and that defined by the $\oa_{\omega l m}$ is the 
``out'' vacuum. One is normally interested in the production of particles on $\scrip$, {\it i.e.,}
in the quantity $\langle 0 | {\overline N}_{\omega l m} | 0\rangle$ as $\tV \rightarrow \infty$, 
where ${\overline N}_{\omega l m} = \oa^\dagger_{\omega l m} \oa_{\omega l m}$ is the number 
operator in the ``out'' vacuum. It is then easily shown that this quantity is determined by the 
second of the two Bogoliubov coefficients,
$$\eqalign{\alpha_{\omega\omega'}~~ &=~~ \langle f_w, \of_{w'}\rangle\cr \beta_{\omega \omega'}~~ 
&=~~ -~ \langle f_\omega, \of^*_{\omega'}\rangle,\cr}\eqno(2.6)$$
that relate the two descriptions of the quantum field in (2.5), according to
$$\langle 0 | {\overline N}_\omega | 0\rangle~~ =~~ \int_0^\infty d\omega' | \beta(\omega' 
\omega)|^2.\eqno(2.7)$$
We have suppressed the dependence on $l,m$ because the geometric optics approximation that will be 
used in this paper is invalid for higher angular momentum modes. However, these modes are expected 
to contribute little either to the flux or to the spectrum of the radiation because of the 
centrifugal potential, which would cause them to scatter to infinity before they encounter the 
region of high curvature(see, for example, B. S. DeWitt in ref.[4]) In the ``out'' region,
$$\eqalign{f_{\omega}~~ &\approx~~ {1 \over {\sqrt{4\pi\omega}}}e^{-i\omega G(\tU)}\cr
\of_{\omega}~~ &\approx~~ {1 \over {\sqrt{4\pi\omega}}}e^{-i\omega \tU},\cr}\eqno(2.8)$$
giving
$$\beta(\omega' \omega)~~ =~~ {1 \over {2\pi}} \sqrt{\omega \over {\omega'}} \int_{-\infty}^{\tU_o}
d\tU e^{-i\omega \tU} e^{-i {\omega'} G(\tU)},\eqno(2.9)$$
where we have used (2.6), and where $\tU=\tU_o$ represents the last outgoing ray that originated in 
an incoming packet from $\scrim$. An equivalent and alternative expression, which constructs 
$\beta(\omega' \omega)$ on $\scrim$, is
$$\beta(\omega' \omega)~~ =~~ {1 \over {2\pi}} \sqrt{{\omega'} \over \omega} \int_{-\infty}^{\tV_o}
d\tV e^{-i{\omega'} \tV} e^{-i \omega F(\tV)},\eqno(2.10)$$
using the asymptotic forms
$$\eqalign{f_{\omega}~~ &\approx~~ {1 \over {\sqrt{4\pi\omega}}}e^{-i\omega \tV}\cr
\of_{\omega}~~ &\approx~~ {1 \over {\sqrt{4\pi\omega}}}e^{-i\omega F(\tV)}\cr}\eqno(2.11)$$
on $\scrim$, and where $\tV=\tV_o$ is the last incoming ray that turns into an outgoing packet 
on $\scrip$.

One may now compute the components of the stress energy tensor of the scalar field from the 
usual expression
$$\langle 0 | T_{\mu\nu}(x) |0\rangle~~ =~~ \lim_{x'\rightarrow x} {\cal D}_{\mu\nu'} {1 \over 2}
G^{(1)}(x,x'),\eqno(2.12)$$
where $G^{(1)}(x,x')$ is Hadamard's elementary function,
$$G^{(1)}(x,x')~~ =~~ \langle 0 | \{\phi(x),\phi(x')\} |0\rangle, \eqno(2.13)$$
and ${\cal D}_{\mu\nu'}(x,x')$ is a non-local operator defined by the form of the stress-energy 
tensor and point-splitting. The expression in (2.12) must obviously be regularized and then
renormalized, upon which one obtains, in particular, an expression for the radiated flux on 
$\scrip$,${}^{[4]}$
$$T_{\tU\tU}(\tV)~~ =~~  {1 \over {24\pi}} \left[{{F'''} \over {{F'}^3}} - {3\over 2}\left({{F''} 
\over {{F'}^2}}\right)^2\right],\eqno(2.14)$$
and a corresponding relation in terms of $G(\tU)$.

It is more interesting, however, to recover the radiated flux, (2.14), from a consistency condition: 
the integrated flux over $\scrip$ must equal the total radiated energy as calculated by integrating 
the radiation spectrum over all frequencies, {\it i.e.,}
$$\int_{-\infty}^{\tU_o} d\tU \langle 0| T_{\tU\tU}|0\rangle~~ =~~ \int_0^\infty d\omega \omega \langle 0|
{\overline N}(\omega) |0\rangle~~ =~~ \int_0^\infty d\omega \omega \int_0^\infty d\omega' 
\beta^*(\omega'\omega) \beta(\omega'\omega). \eqno(2.15)$$
Using $\tU=F(\tV)$, we can transform the integral over future null infinity to one over past null 
infinity, and write
$$\eqalign{\int_{-\infty}^{\tV_o} &d\tV F'(\tV) \langle 0| T_{\tU\tU} |0\rangle~~ =~~ \int_0^\infty 
d\omega \omega \langle 0|{\overline N}(\omega) |0\rangle\cr &=~~ {1 \over {4\pi^2}} 
\int_{-\infty}^{\tV_o} d\tV \int_{-\infty}^{\tV_o} d\tV' \int_0^\infty d\omega \int_0^\infty d\omega' 
\omega' e^{i{\omega'}(\tV'-\tV)} e^{i\omega[F(\tV')-F(\tV)]}~ .\cr}\eqno(2.16)$$
While it may seem rather indirect, our reason for computing the radiated flux in this way is 
the following. When the singularity is globally naked, the Cauchy horizon will intersect future 
null infinity in the retarded past of the apparent horizon, at some point, say $\tU_o$. This 
means that the integration in expression (2.9) for $\beta(\omega'\omega)$ will not extend over all of 
$\scrip$, as it does for the black hole, but only the portion of $\scrip$ that is in the retarded past 
of $\tU=\tU_o$. It is because, as we have pointed out in the introduction, any ray originating at such 
a value of $\tV$ on $\scrim$ as would translate into an outgoing ray in the future of $\tU=\tU_o$ would 
never arrive at $\scrip$, being, instead, absorbed by the singularity. If (2.16), for 
any $\tV_o$, reproduces the correct expression, (2.14), for the radiated flux as computed by a direct 
application of (2.12), it increases our confidence in the spectrum obtained from (2.7) even when 
the singularity is globally naked. $\tV_o$ represents the last null ray that originates on $\scrim$ 
and is able to reach $\scrip$. Both sides of equation (2.15) are infinite, but this does not 
trouble us as we are interested only in obtaining the flux. The infinite result is because, when the back 
reaction of spacetime is not accounted for, particle production will occur indefinitely even 
though the system loses energy. Energy conservation requires, therefore, that the back reaction 
will dominate at some stage. Before this stage is reached, however, (2.16) should serve as a 
good approximation of the actual physical situation. 

The integrals over $\omega$ and $\omega'$, on the right hand side of (2.16), can be performed
to yield,
$$\int_{-\infty}^{\tV_o} d\tV F'(\tV) \langle 0| T_{\tU\tU} |0\rangle~~ =~~ {{-i} \over {4\pi^2}} 
\int_{-\infty}^{\tV_o} d\tV \int_{-\infty}^{\tV_o} d\tV' {1 \over {(\tV'-\tV)^2 [F(\tV')-F(\tV)]}}.
\eqno(2.17)$$
The integral on the right has a pole at $\tV'=\tV$ and contributions from points $\tV'\neq \tV$ vanish 
identically (as is seen by interchaging $\tV$ and $\tV'$ in (2.17)). Calling $z=\tV-\tV'$, the r.h.s. 
of (2.17) becomes 
$${{-i} \over {4\pi^2}}\int_{-\infty}^{\tV_o} d\tV \int_{\tV-\tV_o}^\infty dz {1 \over {z^2[F(\tV-z)-
F(\tV)]}}. \eqno(2.18)$$
We will define the $z$ integral as the contribution from the portion of the contour, in the 
complex $z-$plane, shown in figure 2, that runs  from $\tV-\tV_o$ to $-\epsilon$ along the 
real line (II), along the infinitesimal semi-circle, $C_o$, of radius $\epsilon$ around the 
origin in the upper half plane, and from $+\epsilon$ to infinity along the real line (III). The
non-vanishing contribution from the $z-$integral in (2.18) is then just $-\pi i b_{-1}(\tV)$ where
where $b_{-1}(\tV)$ is the residue of the integrand at $z=0$. 
\vskip 0.1in
\centerline{\epsfysize=1.5in \epsfbox{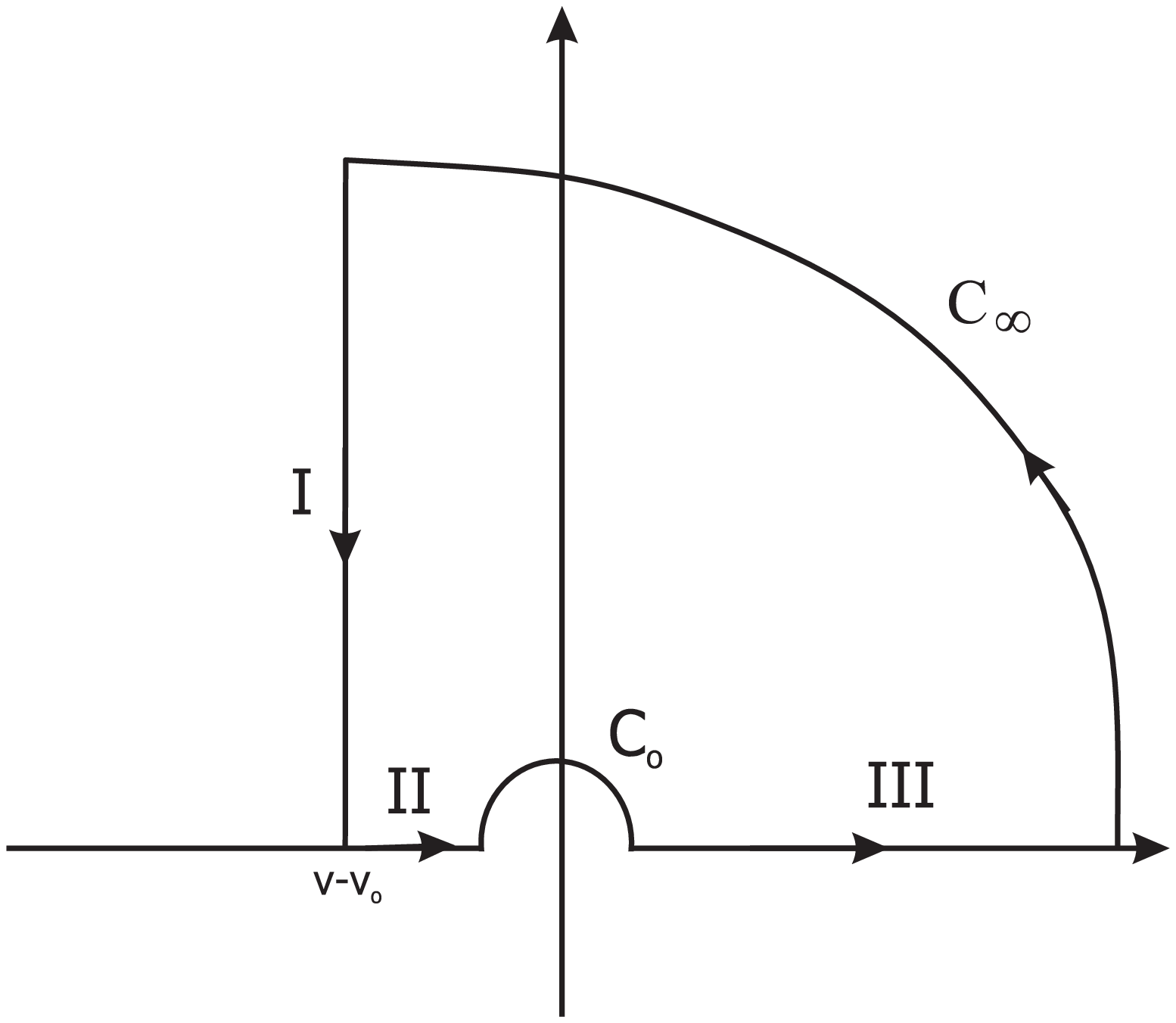}}
\centerline{\Tenpoint Figure 2. The Contour $C$}
\vskip 0.1in

\noindent This is easily seen as follows: the value of the contour integral is identically 
zero, {\it i.e.,} $I+II+III+C_o=0$, as is the contribution from the semi-circle at infinity, 
$C_\infty$. Thus we may write the integral in (2.18) as 
$$\int_{-\infty}^{\tV_o} d\tV \int_{\tV-\tV_o}^\infty dz {1 \over {z^2[F(\tV-z)-F(\tV)]}}~~ =~~ 
- \int_{-\infty}^{\tV_o} d\tV \int_I dz {1 \over {z^2[F(\tV-z)-F(\tV)]}}. \eqno(2.19)$$
Next, consider the contour shown in figure 3. 
\vskip 0.1in
\centerline{\epsfysize=1.5in \epsfbox{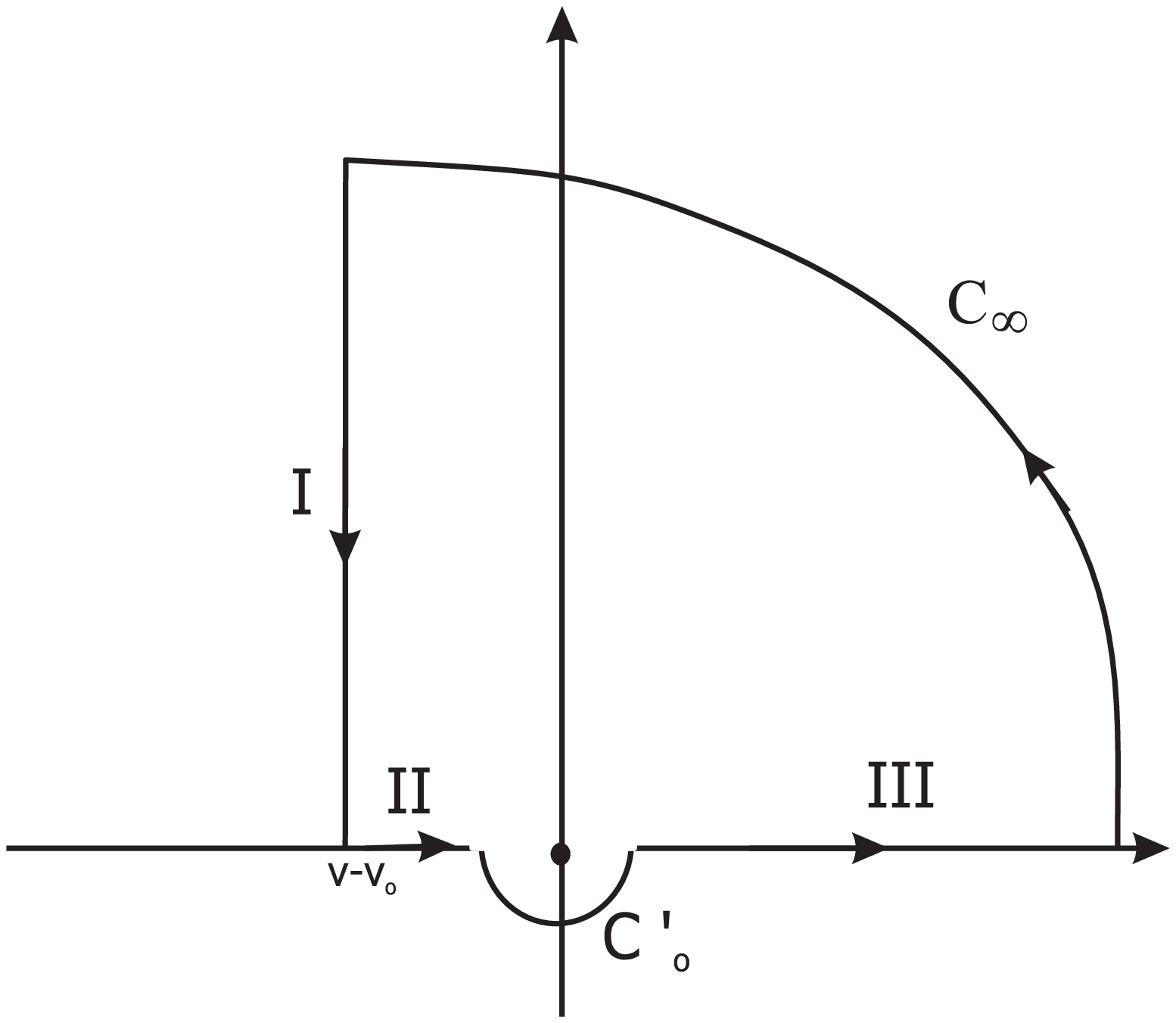}}
\centerline{\Tenpoint Figure 3. The Contour $C'$}
\vskip 0.1in

\noindent The value of this contour integral is $2\pi i b_{-1}(\tV)$, 
{\it i.e.,} $I+II+III+{C'}_o = 2\pi i b_{-1}(\tV)$. 
Combining these two results, one finds $I = \pi i b_{-1}(\tV) - (II+III)$. However, because 
$$\int_{-\infty}^{\tV_o} d\tV \int_{II+III} dz {1 \over {z^2[F(\tV-z)-F(\tV)]}}~~ =~~ 0,\eqno(2.20)$$
it follows immediately that value of the integral, as defined above, is
$${{-i}\over {4\pi^2}} \int_{-\infty}^{\tV_o} d\tV \int_{\tV-\tV_o}^\infty dz {1 \over {z^2
[F(\tV-z)-F(\tV)]}}~~ =~~ -{1 \over {4\pi}} \int_{-\infty}^{\tV_o} d\tV b_{-1}(\tV). \eqno(2.21)$$
The residue can be evaluated by expanding the integrand about $z=0$. We have
$${1 \over {z^2[F(\tV-z)-F(\tV)]}}~~ =~~ {1 \over {z^3}} \left[1 + {z \over {2!}}{{F''} \over {F'}}
- {{z^2} \over {3!}}{{F'''} \over {F'}} + {{z^2} \over {(2!)^2}}\left({{F''} \over {F'}}\right)^2
+~...~ \right],\eqno(2.22)$$
giving
$$b_{-1}(\tV)~~ =~~ {1 \over {6F'}} \left[{3\over 2}\left({{F''} \over {F'}}\right)^2 - {{F'''}
\over {F'}}\right],\eqno(2.23)$$
or, inserting (2.23) into (2.21) and using (2.17), precisely the result in (2.14). The flux 
was originally obtained from (2.12) using standard point-splitting techniques. We have  
recovered it directly from a consistency condition, independently of details of the collapse.
\vskip 0.25in

\noindent{\bf 3. The Collapse of Inhomogeneous Dust.}

We will apply the expressions obtained in (2.9) and (2.23) for the radiation spectrum and 
flux respectively, to the marginally bound, self-similar collapse of inhomogeneous dust. 
Although this model has been examined in detail by us elsewhere,${}^{[6,7]}$ we include here a 
brief analysis of the causal structure of the spacetime, both for the sake of completeness
as well as to set our notation for the succeeding sections. It is described by the stress 
energy tensor
$$T_{\mu\nu}~~ =~~ \epsilon(t,r) \delta^0_\mu \delta^0_\nu.\eqno(3.1)$$
The metric is well known and given in comoving coordinates${}^{[10]}$ by
$$ds^2~~ =~~ dt^2~ -~ \tR^{'2}(t,r)  dr^2~ -~ \tR^2(t,r) d\Omega^2, \eqno(3.2)$$
where the dust cloud is thought of as made up of concentric shells, each labeled by $r$.
$\tR'(t,r)$ is the derivative of $\tR(t,r)$ with respect to $r$ and $\tR(t,r)$ is the physical
radius (the area of a shell labelled by $r$ is $4\pi R^2(t,r)$) obeying, in the particular
case of the marginally bound self similar collapse,
$$\tR(t,r)~~ =~~ r \left[1~ -~ {{3\sqrt{\lambda}} \over 2} {t \over r} \right]^{2/3}.
\eqno(3.3)$$
The physical radius is seen to depend on one parameter, $\lambda$, (the ``mass parameter'').
This parameter determines the total mass, $M(r)$, lying within the shell labeled by $r$ as
$2GM(r) = \lambda r$. The total mass of the dust is therefore $2GM = \kappa = \lambda r_o$
where $r_o$ labels the outer boundary of the cloud. Now it can be shown that $\tR(t,r) = 0$ is
a curvature singularity. This means that the singularity curves are
$t_o(r) = 2r/ (3\sqrt{\lambda})$, so that the last shell becomes singular at the time
$t_o = 2/3 \sqrt{r_o^3/ \kappa}$.

Beyond $r=r_o$ spacetime is described by the the Schwarzschild solution
$$ds^2~~ =~~ \left(1 - {{\kappa} \over R}\right) dT^2~ -~ \left( 1 -
{{\kappa} \over R}\right)^{-1} dR^2~ -~ R^2 d\Omega^2 \eqno(3.4)$$
and the first and second fundamental forms of the two patches must be 
matched at the boundary. This has been done in the past and gives
$$\eqalign{T_o(t)~~ &=~~ -~ 2{\sqrt{\kappa\tR_o}}~ -~ {2 \over 3} \tR_o {\sqrt{
{\tR_o} \over {\kappa}}}~ +~ \kappa \ln |{{{\sqrt{\tR_o}} + {\sqrt{\kappa}}} \over
{{\sqrt{\tR_o}} - {\sqrt{\kappa}}}}|\cr &=~~ t~ -~ {2 \over {3\sqrt
{\kappa}}}r_o^{3/2}~ -~ 2 \sqrt{{\kappa \tR_o}}~ +~ \kappa \ln |{{{\sqrt{\tR_o}} +
{\sqrt{\kappa}}} \over {{\sqrt{\tR_o}} - {\sqrt{\kappa}}}}|\cr R_o(t)~~ &=~~ r_o \left[
1 - a {t \over r_o} \right]^{2/3},\cr} \eqno(3.5)$$
where we have set $a = 3\sqrt{\lambda}/2$.

For the marginally bound, self-similar collapse under consideration, it is relatively
simple to find null coordinates for this system. Consider the effective two dimensional
metric,
$$ds^2~~ =~~ dt^2~ -~ \tR^{'2}(t,r) dr^2, \eqno(3.6)$$
and change variables to $z,x$ where $z~ =~ \ln r$, $x=t/r$. This gives
$$\eqalign{ds^2~~ &=~~ r^2 \left[dx^2~ +~ 2xdxdz~ +~ (x^2~ -\tR^{'2}(x)) dz^2\right]\cr
&=~~ r^2 (x^2 - \tR^{'2}) (d\tau^2~ -~ d\chi^2),\cr}\eqno(3.7)$$
where
$$\eqalign{\tau~~ =~~ z~ +~ {1 \over 2} (I_- + I_+)\cr \chi~~ =~~ {1 \over 2}
(I_- - I_+),\cr}\eqno(3.8)$$
in terms of
$$I_\pm (x)~~ =~~ \int {{dx} \over {x\pm\tR'}}.\eqno(3.9)$$
We would like to choose null coordinates such that in the limit as $\lambda
\rightarrow 0$ these reduce to the standard null coordinates in Minkowski
space. Such coordinates are given by
$$\eqalign{u~~ &=~~ \left\{\matrix{+r e^{I_-}&~~~ x-\tR' > 0\cr-r e^{I-}&~~~ x-
\tR' < 0}\right.\cr v~~ &=~~ \left\{\matrix{+r e^{I_+}&~~~ x+\tR' > 0\cr-r e^{I_+}
&~~~ x+\tR'<0}\right.\cr}\eqno(3.10)$$
To further analyze the causal structure, it is now convenient to go over to the
variable $y$ defined by $y = \sqrt{\tR/r}$. In terms of $y$, the integrals $I_\pm$ can be 
written as
$$I_\pm~~ =~~ 9 \int {{y^3 dy} \over {3y^4~ \mp~ ay^3~ -~ 3y~ \mp~
2a}} \eqno(3.11)$$
and the coordinates (3.10) become
$$\eqalign{u~~ &=~~ \left\{\matrix{+r e^{I_-}&~~~ f_-(y) < 0\cr-r e^{I-}&~~~ f_-(y)
> 0}\right.\cr v~~ &=~~ \left\{\matrix{+r e^{I_+}&~~~ f_+(y) < 0\cr-r e^{I_+}
&~~~ f_+(y) > 0}\right. \cr}\eqno(3.12)$$
where
$$f_\pm (y)~~ =~~ 3y^4~ \mp~ ay^3~ -~ 3y~ \mp~ 2a. \eqno(3.13)$$
Let $\alpha^\pm_i$ be the roots of $f_\pm(y)$, for $i~ \epsilon~ \{1,2,3,4\}$. As $f_\pm$
are both real, they admit either 0, 2, or 4 real roots. The integrals can now be put in the form
$$I_\pm~~ =~~ 3 \int dy \left[\sum_{i=1}^4 {{A^\pm_i} \over {(y-\alpha^\pm_i)}}\right],
\eqno(3.14)$$
where the $A^\pm_i$ are constants related to the coefficients of $f_\pm(y)$ and their
roots by,
$$A_i^\pm~~ =~~ {{\alpha_i^{\pm 3}} \over {f'_\pm(\alpha_i^\pm)}}.\eqno(3.15)$$
In particular, the $A^\pm_i$ satisfy $\sum_i A^\pm_i = 1$. If all the roots are real,
the solution is explicitly given by
$$\eqalign{u(y)~~ &=~~ \pm~ r \prod_{i=1}^4 |y~ -~ \alpha^-_i|^{3A^-_i}\cr
v(y)~~ &=~~ \pm~ r \prod_{i=1}^4 |y~ -~ \alpha^+_i|^{3A^+_i}.\cr}\eqno(3.16)$$
We will now consider the case in which there are two real roots and a conjugate
pair of complex roots. As we will shortly show at least two real roots (possibly degenerate)
are required for the existence of a globally naked singularity at the origin so we do not
consider the case when all the roots are complex even though it may be carried out in
the same spirit. Let us order the roots so that the first two, $\alpha_{1,2}$, are a
complex conjugate pair and $\alpha_{3,4}$ are real. From (3.15) it follows that
$A_{1,2}$ is also a complex pair whereas $A_{3,4}$ are real. Then the integrals
are of the form
$$\eqalign{I~~ =~~ 3 \int dy &\left[\sum_{i=1}^4 {{A_i} \over {(y-\alpha_i)}}\right]\cr 
&=~~ 3 \left[A \ln (y - \alpha)~ +~ A^* \ln (y - \alpha^*)~ +~ \sum_{i=3,4} A_i \ln |
y - \alpha_i|\right],\cr}\eqno(3.17)$$
where $\alpha, \alpha^*$ are the complex roots and $A,A^*$ are the (complex)
coefficients. Putting
$$A~~ =~~ |A| e^{i\phi},~~~~~~~~~~ y-\alpha~~ =~~ |y-\alpha| e^{i\xi}, \eqno(3.18)$$
so that the $u,v$ coordinates have the explicit (and formal) solution
$$\eqalign{u(y)~~ &=~~ \pm r|y-\alpha^-|^{6|A^-|\cos \phi^-}e^{-6|A^-|\xi^-\sin\phi^-}
\Pi_{i=3,4} |y-\alpha^-_i|^{3A^-_i}\cr v(y)~~ &=~~ \pm r|y-\alpha^+|^{6|A^+|\cos\phi^+}
e^{-6|A^+|\xi^+\sin\phi^+} \Pi_{i=3,4} |y-\alpha^+_i|^{3A^+_i}.\cr}\eqno(3.19)$$
Consider the center ($r=0$) at early times, $t<0$. Then, because $y = (1-at/r)^{1/3}
\rightarrow \infty$, (3.19) gives (when all roots are real)
$$\eqalign{u~~ &\rightarrow~~ -~ r |y|^{3\sum_i A^-_i}~~ =~~
-r (1~ -~ a{t\over r})~~ \rightarrow at\cr v~~ &\rightarrow~~ -~ r |y|^{3\sum_i A^+_i}~~
=~~ -r (1~ -~ a{t\over r})~~ \rightarrow at. \cr}\eqno(3.20)$$
This line is therefore given by $u = v$. When two of the roots are complex conjugates of 
each other, the line is still $u=v$ as we now show. Note that
$$\xi~~ =~~ \tan^{-1} \left({{{\rm Im} (-\alpha)} \over {{{\rm Re}(y-\alpha)}}}\right)$$
($y$ is real), so that as $y\rightarrow \infty$, $\xi \rightarrow 0$. Then
clearly
$$\eqalign{u~~ &\rightarrow~~ -~ r |y|^{3(2{\rm Re} A^- + A^-_3 + A^-_4)}\cr v~~ &
\rightarrow~~ -~ r |y|^{3(2{\rm Re} A^+ + A^+_3 + A^+_4)}\cr}\eqno(3.21)$$
but, since $\sum_i A^\pm_i = 1$, we have the same result as before.

The general solutions in (3.16) and (3.19) are useful to analyze another limit, namely
the singularity at $r \rightarrow at$. This means that $y \rightarrow 0$. Now when
$y\rightarrow 0$, $f_-(y)>0 $ and $f_+(y)<0$. Then we see that (if all roots are real)
$$\eqalign{u~~ &=~~ -~ r \prod |\alpha^-_i |^{3A^-_i}\cr v~~ &=~~ r \prod |\alpha^+_i
|^{3A^+_i}\cr}\eqno(3.22)$$
and, in particular,
$${v\over u}~~ =~~ -~ c~~ =~~ -~ \prod_i {{|\alpha^+_i |^{3A^+_i}} \over
{|\alpha^-_i |^{3A^-_i}}}, \eqno(3.23)$$
which is a negative constant, in general $\neq -1$. The singularity is therefore {\it
spacelike} until the last shell, $r=r_o$, collapses at $t=t_o = r_o/a$. The case
of a pair of conjugate complex roots trivially gives the same result. Beyond this point
the singularity will be spacelike because it is just the Schwarzschild singularity in
the exterior region. The behavior of the origin, $r=0, t=0$, is peculiar. It is
the meeting point between two lines $u=v$ and $u=-cv$ and its nakedness (coveredness)
is far from clear. However, if a null ray originating at this point reaches the
boundary at Kruskal coordinate $U < 0$ in the Schwarzschild region, it will reach
$\scrip$ and then the origin will be globally naked.

We will be interested in the earliest null ray leaving the singularity and reaching
$\scrip$ (the Cauchy Horizon) as well as the earliest null ray that strikes the singularity
from $\scrim$. These rays can be expected to intersect the first singular
shell at $r=0, t=0$, so it is natural to carefully examine the null rays passing through this
point. The origin, being the intersection of the lines $u=v$ and $v= -cu$ ($c \neq 1$ in general),
corresponds to the point $u=0=v$. Now any null ray traveling toward $\scrip$ with $u=0$ must
have either $r=0$ or $I_- \rightarrow -\infty$. Therefore, when $r \neq 0$, such a ray is possible
if and only if $y=\alpha^-_k$ for some {\it real} root, $\alpha^-_k$ of the polynomial $f_-(y)$. Indeed
such a root may not exist, in which case the singularity is not naked as no null rays can emanate
from it. In this case, a black hole is formed, as shown in figure 3. 
\vskip 0.1in
\centerline{\epsfysize=1.5in \epsfbox{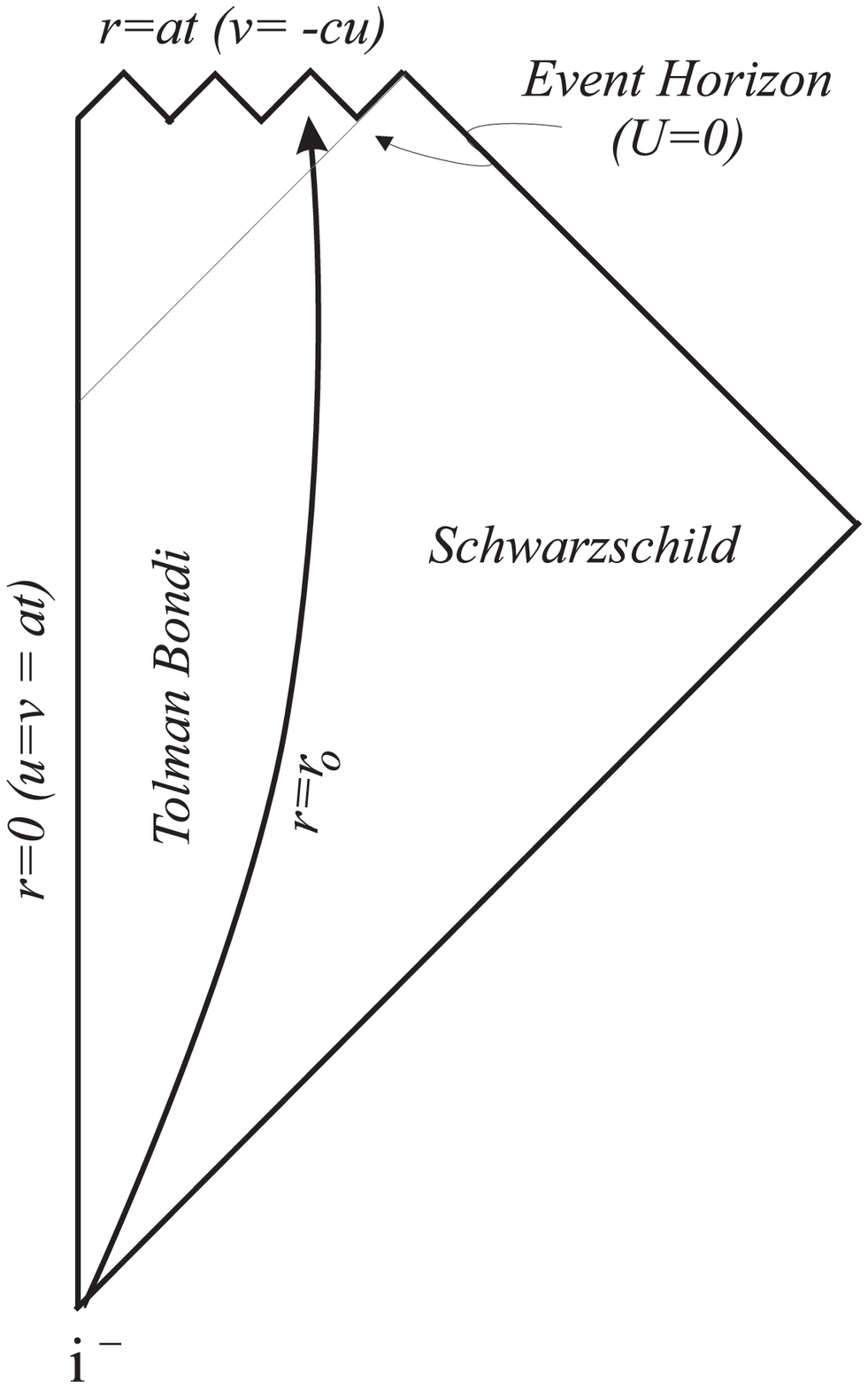}}
\centerline{\Tenpoint Figure 4. Formation of a black hole}
\vskip 0.1in

\noindent If a real root exists however, at least one null ray leaves this point and reaches the
boundary. The existence of real roots of the polynomials $f_-(y)$ is therefore a necessary condition
for the nakedness of the origin. This places a constraint on the possible value of the constant $a$
in the mass function. One finds that real roots exist provided that $a < a_c \sim 0.638$.${}^{[9]}$
Each root corresponds to a null ray emanating from $u=0=v$ and there are at least two of them, if any
at all. Because $y = \alpha_i$ implies that $t = r(1-\alpha_i^3)/a$, we choose the largest real
root of $f_-(y)$ as the one that gives the earliest null ray emanating from $u=0=v$ and call it
$\alpha_-$. Thus, $y=\alpha_-$ is the Cauchy horizon.

A similar reasoning can now be given for the incoming rays passing through $u=0=v$. Again any
ray with $v=0$ for $r\neq 0$ must have $I_+ \rightarrow -\infty$, which is possible only
if $y=\alpha^+_k$ for some {\it real} root, $\alpha^+_k$ of the polynomial $f_+(y)$. Now,
$f_+(y)$ admits two real roots, one unphysical (negative) and one positive.  Again,
call the (positive) physical root $\alpha_+$.

What we have described above is pictured in the Penrose diagram of figure 1. It is to be 
expected that the behavior of quantum fields will be extremely sensitive to 
the collapse scenario being considered, that is to whether the mass parameter $a$ lies below 
or above its critical value, and this is the topic of the next section.
\vskip 0.25in

\noindent{\bf 4. Radiation flux and spectrum for black holes and naked singularities.}

We will henceforth consider rays in the neighborhood of the lines given by $y = \alpha_-$
for outgoing rays and $y = \alpha_+$ for incoming rays. The precise values of $\alpha_\pm$ in terms
of the mass parameter will not interest us for this work but we will Taylor expand about these two
values, considering $y_\pm = \ty_\pm + \alpha_\pm$.

Returning to (3.5), one can rewrite the Schwarzschild radial coordinate and time on the boundary as
follows
$$\eqalign{R_o(y)~~ &=~~ r_o y^2\cr T_o(y)~~ &=~~ -~ {{r_o} \over a} y^3~ -~  {4 \over 3} a r_o y~
-~ {4 \over 9} a^2 r_o \ln |{{3y/2a - 1} \over {3y/2a + 1}}|. \cr}\eqno(4.1)$$
Therefore, the Eddington-Finkelstein null coordinates on the boundary, $\tU_o(y) = T_o(y)-R_{o*}(y)$,
$\tV_o(y) = T_o(y) + R_{o*}(y)$, (where $R_{o*}$ is the tortoise coordinate) take the form
$$\eqalign{\tU_o(y)~~ &=~~ -~ {{r_o} \over a} y^3~ -~  {4 \over 3} a r_o y~ -~ r_o y^2 ~ -~
{8\over 9} a^2 r_o \ln |3y/2a - 1|\cr \tV_o(y)~~ &=~~ -~ {{r_o} \over a} y^3~ -~  {4 \over 3}
a r_o y~ +~ r_o y^2 ~ +~ {8\over 9} a^2 r_o \ln |3y/2a + 1|.\cr}\eqno(4.2)$$
It is now clear that the earliest null outgoing ray, $u=0$, from the origin (the Cauchy Horizon)
within the cloud strikes the boundary at $y=\alpha_-$ and translates into the null outgoing ray
$$\tU_o^{(0)}~~ =~~ -~ {{r_o} \over a} \alpha_-^3~ -~  {4 \over 3} a r_o \alpha_-~ -~ r_o
\alpha_-^2 ~ -~ {8\over 9} a^2 r_o \ln |3 \alpha_-/2a - 1|,\eqno(4.3)$$
which is never infinite ($2a/3$ is not a root of $f_-(y)$). This null ray corresponds to a finite
value of $\tU$ and will therefore reach $\scrip$, so the existence of real roots of $f_-(y)$
turns out to be not just necessary, but a sufficient condition for the origin to be globally naked.
The same argument applies to the infalling ray(s): the earliest null ray to pass
through the origin is the ray corresponding to the value $y=\alpha_+$, or
$$\tV_o^{(0)}~~ =~~ -~ {{r_o} \over a} \alpha_+^3~ -~  {4 \over 3} a r_o \alpha_+~ +~ r_o
\alpha_+^2 ~ +~ {8\over 9} a^2 r_o \ln |3 \alpha_+/2a + 1|\eqno(4.4)$$
and, again, since $-2a/3$ is not a root of $f_+(y)$, $\tV$ is not infinitely negative and such a
ray will have come from $\scrim$. Thus, the existence of a positive real root of $f_+(y)$ is
sufficient to ensure that at least one infalling ray from $\scrim$ will intersect the origin.

The next question we must address is the relationship between the $\tU, \tV$ coordinates in the
exterior and the $u,v$ coordinates (equations (18, 21)) on the boundary. This is difficult to do in
general, but if we confine our study to rays that are ``close'' to $u=0$ and $v=0$ we can arrive at
some conclusion regarding the quantum radiation on $\scrip$ near the Cauchy horizon. ``Close''
will be taken to mean linearizations about $y=\alpha_\pm$ respectively for incoming rays and
outgoing rays.

First consider outgoing rays. For $y \sim \alpha_-$, define $y=\ty+\alpha_-$ and find that for 
small $\ty$
$$I_-~~ \sim~~ \gamma_- \ln \ty~~ +~~ {\cal O}(y),\eqno(4.5)$$
where
$$\gamma_-~~ =~~ {{3\alpha_-^3} \over {f'_-(\alpha_-)}},\eqno(4.6)$$
giving
$$u~~ =~~ -r|\ty|^{\gamma_-}~~ \rightarrow~~ y-\alpha_-~~ =~~ \left(- {u \over r} \right)^{1/
\gamma_-}.\eqno(4.7)$$
Therefore in terms of $u$ (on the boundary) we can write $\tU$ as follows
$$\tU~~ \sim~~ \tU^{(0)}(\alpha_-)~ +~ \Gamma_-(\alpha_-) (y-\alpha_-) ~~ =~~ \tU^{(0)}(
\alpha_-)~ +~ \Gamma_-(\alpha_-) \left(-{{u} \over {r_o}} \right)^{1/\gamma_-},\eqno(4.8)$$
where
$$\Gamma_-~~ =~~ -~9 {{r_o \alpha_-^3} \over {a (3\alpha_- -2a)}}~~ <~~ 0~~ {\rm when}~~ a < a_c .
\eqno(4.9)$$
Likewise, for incoming rays,
put $y = \ty+\alpha_+$ and find that
$$I_+~~ =~~ \gamma_+ \ln \ty~~ +~~ {\cal O}(y),\eqno(4.10)$$
where
$$\gamma_+~~ =~~ {{3\alpha_+^3} \over {f'_+(\alpha_+)}},\eqno(4.11)$$
giving
$$v~~ =~~ -r|\ty|^{\gamma_+}~~ \rightarrow~~ y-\alpha_+~~ =~~ \left(- {v \over r} \right)^{1/
\gamma_+}.\eqno(4.12)$$
Thus, in terms of $v$ (on the boundary) we can write $\tV$ as follows
$$\tV~~ \sim~~ \tV^{(0)}(\alpha_+)~ +~ \Gamma_+ (\alpha_+)(y-\alpha_+)~~ =~~ \tV^{(0)}
(\alpha_+)~ +~ \Gamma_+ (\alpha_+) \left(-{{v} \over {r_o}}\right)^{1/\gamma_+},\eqno(4.13)$$
where
$$\Gamma_+~~ =~~ -~9 {{r_o\alpha_+^3} \over {a (3\alpha_+ + 2a)}}~~ <~~ 0~~  {\rm when}~~  a<a_c .
\eqno(4.14)$$
We are now in a position to compute the radiated power close to the Cauchy horizon in
the geometric optics approximation. consider a ray $\tV=$ const. in the infinite past. We are interested 
only in the region on $\scrip$ that is close to the Cauchy horizon, so the approximations in (4.8) and (4.13) 
will suffice. As the null ray crosses the boundary, we have
$$\tV(v)~~ =~~ \tV^{(0)}~ +~ \Gamma_+ \left(-{v\over {r_o}} \right)^{1 \over {\gamma_+}}.
\eqno(4.15)$$
This expression can be inverted to give
$$v(\tV)~~ =~~ -~ r_o \left[ {{\tV^{0} - \tV} \over {|\Gamma_+|}} \right]^{\gamma_+}, \eqno(4.16)$$
where we have used the fact that $\Gamma_+$ is negative. Next, reflecting about the center (here,
$u=v$) gives
$$u(\tV)~~ =~~ -~ r_o \left[ {{\tV^{0} - \tV} \over {|\Gamma_+|}} \right]^{\gamma_+}. \eqno(4.17)$$
Now as the outgoing ray crosses the outer boundary, we have the relation
$$\eqalign{\tU(u)~~ &=~~ \tU^{(0)}~ -~  \Gamma_- \left(-{u \over {r_o}} \right)^{1 \over {\gamma_-}}
\cr \rightarrow~~ \tU(\tV)~~ &=~~ \tU^{(0)}~ -~ |\Gamma_-|\left[{{\tV^{0} - \tV}\over {|\Gamma_+|}}
\right]^{{\gamma_+} \over {\gamma_-}},\cr} \eqno(4.18)$$
where now we have used the fact that $\Gamma_-$ is negative. Thus, the right hand side of (4.18) is
$F(\tV)$ and it has the form
$$F(\tV)~~ =~~ A~ -~ B (\tV^{(0)} - \tV)^{{\gamma_+} \over {\gamma_-}}, \eqno(4.19)$$
where $B$ is a positive constant which is given in terms of the roots, $\alpha_\pm$ given before.
We can now write down the power radiated as a function of $\tV$ using (2.14),
$$T_{\tU\tU}(\tV)~~ \approx~~ {1 \over {48\pi B^2}} \left[{{1-\gamma^2} \over 
{\gamma^2(\tV^{(0)} - \tV)^{2\gamma}}}\right] ~~~~~ \gamma \neq 0, \eqno(4.20)$$
where
$$\gamma~~  =~~ {{\gamma_+} \over {\gamma_-}}.\eqno(4.21)$$
The result can be written in terms of $\tU$ by inversion,
$$T_{\tU\tU}(\tU)~~ \approx~~ {1 \over {48\pi}} \left[{{1-\gamma^2} \over 
{\gamma^2(\tU^{(0)} - \tU)^{2}}}\right],\eqno(4.22)$$
which diverges as the Cauchy horizon is approached.

Let us now consider the case when the origin is not naked, {\it i.e.}, all roots of the 
polynomial $f_-(y)$ are complex. This means of course that the Cauchy 
horizon is formed in the retarded future of the event horizon as shown in figure 4. 
>From the expression (4.3) for $\tU$, this implies that the event horizon 
intersects the boundary at $y \rightarrow 2a/3$. We will be interested in late 
times, so consider a ray that is close to the event horizon and that therefore 
intersects the boundary at $y = 2a/3 + \ty$. For such a ray, 
$$\tU~~ \sim~~ -~ 4M \ln|\ty|,\eqno(4.23)$$
where $M = 2ar_o^2/9$ is the total mass of the cloud. Continuing backward 
into the cloud, it is necessary to retain only terms that are linear in 
$\ty$ in the expression for $u$. This ray translates 
into the ray
$$u~~ =~~ r_o \exp(I_-({{2a} \over 3} + \ty))~~ =~~ r_o\exp({3 \over a} \ty)~~ 
\sim~~ r_o(1+{{3\ty}\over a}),\eqno(4.24)$$
or
$$\ty~~ \sim~~ {a\over 3}\left({u \over {r_o}} - 1\right),\eqno(4.25)$$
which, when substituted back into the expression for $\tU$ in (10), gives 
$\tU$ as a function of $u$ inside the cloud:
$$\tU~~ =~~ -4M \ln |{a \over 3}\left({u \over {r_o}} - 1\right)|.\eqno(4.26)$$
Reflecting at the center ($v=u$), in terms of the advanced coordinate, 
$v$, within the cloud as
$$\tU~~ =~~ -4M \ln |{a \over 3}\left({v \over {r_o}} - 1\right)|.\eqno(4.27)$$
We must now find a relationship between $v$ and $\tV$ in the exterior 
region. Let us suppose that the ray $\tU = \infty$ traced backward 
to the ray $\tV = \tV_o$. It is not important to know the precise value 
of $\tV_o$ though this can be done and gives complicated expressions 
in terms of the roots of the polynomials $f_\pm(y)$. It is clear that 
this value, $\tV_o$ of $\tV$ corresponds to some given value $y_o$ of 
$y$ on the boundary. Consider a linearization about this value $y_o$ 
($y = y_o+\ty$) so that
$$\tV~~ =~~ \tV_o~ +~ \tV'(y_o)\ty~ +~ ...\eqno(4.28)$$
where $\tV'(y)$ is the derivative of $\tV$ w.r.t. $y$. Again, expanding 
$v$ about this value, $y_o$ of $y$ on the boundary $r=r_o$ gives
$$v~~ =~~ r_o\exp(I_+(y_o+\ty))~~ =~~ v_o~ +~ v'(y_o)\ty~ +~ ... \eqno(4.29)$$
The precise values of $\tV_o, v_o, \tV'(y_o)$ and $v'(y_o)$ will not 
interest us for the following analysis. What is important is that 
$$v~~ =~~ v_o~ +~ {{v'(y_o)} \over {\tV'(y_o)}}(\tV - \tV_o)\eqno(4.30)$$ 
is linear to the order of interest, and that $v_o = r_o$ (from (4.29)), so that
$$\tU~~ =~~ -4M \ln|{{\tV - \tV_o}\over B}|~~ =~~ F(\tV),\eqno(4.31)$$
where $B= 3r_o\tV'(y_o)/av'(y_o)$ is an irrelevant constant. Applying (2.14), it 
follows that 
$$T_{\tU\tU}(\tU)~~ \approx~~ {1 \over {192\pi M^2}}\eqno(4.32)$$
to leading order. The radiation flux is seen to approach a constant as the 
horizon is approached.

The marginally naked singularity does not exist in this model. The singularity 
would be marginally naked if the Cauchy horizon coincided with the event horizon 
for some value of $a$. However, we have seen that the event horizon is given by 
$y=2a/3$, which is not a root of the polynomial $f_-(y)$ (except when $a=0$).
The singularity is therefore either naked ($a\leq a_c$) or covered ($a > a_c$), 
but never marginal.

We now turn to the spectrum of the radiation emitted by the singularities.
The famous black body radiation spectrum of the black hole (see Hawking in ref.[4])
is a direct consequence of the form of $F(\tV)$, given in (4.31), and the fact that 
the integral in (2.9) extends over all of $\scrip$, so that all the outgoing basis 
states are sampled by the wavepackets formed from incoming plane waves that have 
scattered through the spacetime. 

When a naked singularity is formed, the scattering can occur arbitrarily close to 
the singularity, so that $F(\tV)$ (as given by (4.19)) has a significantly different 
dependence on the advanced coordinate $\tV$. But there is another and more 
important difference. If we consider the possibility that a collapsing star does not 
eliminate all of the spacetime to its future, but that the spacetime continues as the 
analytic extension of the spacetime in the past of the Cauchy horizon and that $\scrip$ 
continues to be well defined in its retarded future (and is therefore complete), then 
since no outgoing wavepackets formed from infalling plane waves are able to reach 
$\scrip$ in the retarded future of the Cauchy horizon we see that not all of the 
outgoing basis states on $\scrip$ are sampled by the outgoing wavepackets. A direct 
consequence of this is that the spectrum is non-thermal. A simple calculation of the 
Bogoliubov coefficients, with $F(\tV)$ given in (4.19), yields,${}^{[8]}$
$$|\beta(\w',\w)|^2~~ =~~ {1 \over {4\pi^2\w\w'}} |\sum_{k=0}^\infty
{{(i \w' (B\w)^{-1/\gamma} e^{i\pi/2\gamma})^k} \over {k!}}  \Gamma ({k\over\gamma}
+1)|^2,\eqno(4.33)$$
or
$$|\beta(\w',\w)|^2~~ =~~ {{\w B} \over {4\pi^2  \gamma \w'^{\gamma+1}}}
|\sum_{k=0}^\infty {({i\w B\w'^{-\gamma}e^{-i\pi\gamma /2})^k} \over
{k!}} \Gamma(k\gamma+1)|^2.\eqno(4.34)$$
The first expression, eq. (4.33) above, is useful to analyze the high frequency limit 
($\w'(B\w)^{-1/\gamma} \rightarrow 0)$ limit of the spectrum, for in this limit it is
sufficient to consider only the first term in the series. Integration over $\w'$ then
yields the familiar logarithmic divergence and the spectrum is seen to fall of as 
$1/\w$ in the high frequency region. The second expression, eq. (4.34), serves to 
analyze its low frequency ($\w'(B \w)^{-1/\gamma} \rightarrow \infty$)
behavior. Integration over $\w'$ in this limit shows a power law divergence in the
infrared. This divergence is associated with the fact that there are an infinite number
of quanta in each mode on $\scrip$. The difference between the divergence in the low 
and high frequency regimes may be associated with the red-shifting of modes in the 
proximity of the putative Cauchy horizon. Nevertheless, $|\beta(\w',\w)|^2$ is seen 
to be well behaved as a function of $\w$, falling as $\w$ when $\w \rightarrow 0$.

As mentioned in the introduction, it is possible that the Cauchy horizon should be 
regarded as the natural end point of spacetime, so that the analytical continuation 
we have considered above is not physically acceptable. In this case, $\scrip$ is 
not complete and $\tU$ is no longer a good asymptotic coordinate.

Referring back to figure 1, we see that the transformation to asymptotically flat 
coordinates must take the form
$$U~~ =~~ -2\kappa e^{-\tU/2\kappa}~~ =~~ -2\kappa e^{-\tu/2\kappa}~ +~ U^{(0)},
\eqno(4.35)$$
which defines the asymptotic null coordinate $\tu$ and where $U^{(0)}$ is defined by
$$U^{(0)}~~ =~~  -2\kappa e^{-\tU^{(0)}/2\kappa}, \eqno(4.36)$$
in terms of $\tU^{(0)}$ given earlier. Clearly, its definition is such 
that $\tu$ ranges from $-\infty$ to $+\infty$, while $U$ ranges from $-\infty$ to 
the Cauchy horizon, $U^{(0)}$. A complete outgoing basis set will be defined w.r.t. 
$\tu$ instead of $\tU$. Putting $U = -2\kappa e^{\tU/2\kappa}$ in (4.35) we find 
$$\eqalign{\tu~~ &=~~ \tU^{(0)}~ -~ 2\kappa \ln | 1 - e^{-(\tU-\tU^{(0)})/2\kappa}|~~ 
\sim~~ \tU^{(0)}~ -~ 2\kappa \ln | {{\tU - \tU^{(0)}} \over {2\kappa}} | \cr &\rightarrow~~ 
\tU~ -~ \tU^{(0)}~~ =~~ 2\kappa e^{-(\tu - \tU^{(0)})/ 2\kappa} \cr} \eqno(4.37)$$
(using the fact that we are near the Cauchy horizon). Now relating $\tV$ and $\tu$
(instead of $\tU$ as we did earlier) we find
$$\tu~~ =~~ F(\tV)~~ =~~ \tU_o~ -~ {{2\kappa}\over\gamma} \ln|{{\tV_o-\tV}\over 
{B'}}|, \eqno(4.38)$$
where $B'$ is an irrelevant constant. However, this is precisely the relationship 
between the infalling and outgoing coordinates for a black hole, given in (4.31). 
It will consequently yield thermal radiation,
$$|\beta(\w',\w)|^2~~ =~~ {\kappa\over {\pi\w'}}{1 \over {e^{4\pi\kappa\w/
\gamma} - 1}},\eqno(4.39)$$
at the modified temperature, $T~~ =~~ \gamma/4\pi\kappa$.
This situation is analogous to the marginally naked singularity treated by Hiscock et.
al.${}^{[6]}$
\vskip 0.25in

\noindent{\bf 5. Discussion}

In this paper, we have used the marginally bound, spherically symmetric collapse of 
inhomogeneous, pressureless dust, which admits both classical black hole and naked 
singularity end states, to illustrate some key differences between the Hawking 
radiation from these objects. The central distinguishing feature appears to be the 
rapid flux of radiation that will be emitted from the naked singularity in the approach 
to the Cauchy horizon. The intensity of the radiation is clearly a consequence of the large 
curvatures that are encountered by infalling rays on their way out to future null infinity, 
so we expect this feature to hold true generically, and whenever regions of high curvature 
are visible to the asymptotic observer. In this spirit, the naked singularity may be thought 
of as a region of high curvature that is visible from future null infinity and not necessarily 
as a true singularity of spacetime. This rapid evaporation signals an instability of the 
Cauchy horizon, and may cause the collapsing star to evolve in such a way as to avoid its 
actual formation. It may be the mechanism by which nature avoids naked singularities, in 
which case the Cosmic Censorship hypothesis would originate in the quantum theory. We have 
not addressed the manner in which the appearance of the Cauchy horizon may be avoided as 
this requires a detailed study of the back reaction of spacetime. If nature does in fact 
employ the quantum theory to avoid naked singularities, the magnified luminosity is likely 
to be observable and may allow for a glimpse into the behavior of matter fields in strongly 
curved backgrounds. Again, if this possibility is taken seriously, it becomes necessary to 
look for additional features of the radiation that would distinguish naked singularities 
from other radiating objects. The spectrum of the radiation is one possibility. Although 
it is to be expected that the electromagnetic spectrum reaching the distant observer will 
not be characteristic of the collapsing star but rather of the thermalized debris surrounding 
it, the spectrum radiated in the form of neutrinos and gravitational waves should escape 
the surrounding matter relatively undisturbed.

A central issue in the calculation of the spectrum of the Hawking radiation from spacetimes 
that admit Cauchy horizons, is that of the existence of a complete future null infinity. 
We know of no way to address this question within the semi-classical approach. There are, 
however, only two logical possibilities: {\it (i)} either the spacetime continues beyond 
the Cauchy horizon or {\it (ii)} it terminates at the Cauchy horizon. 

If the local collapse of matter does not destroy the entire universe in its future (it 
is difficult to imagine that it would), the spacetime can be analytically continued beyond 
the Cauchy horizon and a complete future null infinity exists. Yet, no incoming waves on 
past null infinity are able to form wave packets to the future of the Cauchy horizon, 
therefore only a part of $\scrip$ is actually probed. This results in a non-thermal spectrum.
However, it also raises the issue of the consistency of the spectrum derived from the 
Bogoliubov coefficients. We have addressed this issue by showing explicitly that the total 
energy radiated, as computed from the integrated spectrum (derived from the Bogoliubov 
coefficients) is identical to the total energy radiated, as computed from the stress energy 
tensor to leading order. On the other hand, if the local collapse of a star indeed would 
destroy the universe in the future of the Cauchy horizon, so that the spacetime must 
terminate there, we showed that the spectrum of the radiation emitted is then necessarily 
thermal at a modified temperature.

Only a complete theory of quantum gravity can answer the question of the existence of a complete 
$\scrip$ as this depends on the final fate of the collapse. We note, however, that the instability 
of the Cauchy horizon appears to signal that all of $\scrip$ will survive and therefore that the 
spectrum will be non-thermal.
\vskip 0.25in

\noindent{\bf Acknowledgements:}

\noindent We acknowledge the partial support of the {\it Funda\c{c}\~ao para a Ci\^encia e a 
Tecnologia} (FCT), Portugal, under contract number 
CERN/S/FAE/1172/97.

\vfill\eject
\noindent{\bf References}

{\item{[1]}}P. Yodzis, H.-J Seifert and H. M\"uller zum Hagen, Commun. Math. Phys. {\bf 34} (1973) 135;
Commun. Math. Phys. {\bf 37} (1974) 29; D. M. Eardley and L. Smarr, Phys. Rev. D {\bf 19}, (1979) 2239;
D. Christodoulou, Commun. Math. Phys. {\bf 93} (1984) 171; R. P. A. C. Newman, Class. Quantum Grav.
{\bf 3} (1986) 527; B. Waugh and K. Lake, Phys. Rev. D {\bf 38} (1988) 1315; V. Gorini, G. Grillo and
M. Pelizza, Phys. Lett. A {\bf 135} (1989) 154; G. Grillo, Class. Quantum Grav. {\bf 8} (1991) 739; R.
N. Henriksen and K. Patel, Gen. Rel. Gravn. {\bf 23} (1991) 527; I. H. Dwivedi and S. Dixit, Prog.
Theor. Phys. {\bf 85} (1991) 433; P. S. Joshi and I.H. Dwivedi, Phys. Rev {\bf D47} (1993) 5357; I. H.
Dwivedi and P.S. Joshi, Comm. Math. Phys. {\bf 166} (1994) 117; S. Jhingan, P.S. Joshi, T. P. Singh,
Class. Quant. Grav. {\bf 13} (1996) 3057; I. H. Dwivedi, P. S. Joshi, Class. Quant. Grav. {\bf 14}
(1997) 1223; P. S. Joshi and I. H. Dwivedi, Class. Quant. Grav. {\bf 16} (1999) 41; S. Barve, T.P. 
Singh, Cenalo Vaz and L. Witten, Class. Quant. Grav. {\bf 16} (1999) 1727; S. S. Deshingkar, I. H. 
Dwivedi and P. S. Joshi, Phys. Rev. {\bf D59} (1999) 044018; {\it ibid} Gen. Rel. Grav. {\bf 30} (1998) 
1477; S. Jhingan and G. Magli, gr-qc/9902041; {\it ibid} gr-qc/9903103; 

{\item{[2]}}P. S. Joshi, {\it Global Aspects in Gravitation and Cosmology}, Clarendon Press, Oxford,
(1993).

{\item{[3]}}R. Penrose, Riv. Nuovo Cimento {\bf 1} (1969) 252; in {\it General Relativity, An Einstein
Centenary Survey}, ed. S. W. Hawking and W. Israel, Cambridge Univ. Press, Cambridge, London (1979)
581. In its original form, the Cosmic Censorship Hypothesis (CCH) essentially states that: {\it
no physically realistic collapse, evolving from a well posed initial data set and satisfying the
dominant energy condition, results in a singularity in the causal past of null infinity}. There is
also a strong version of the CCH which states that: {\it no physically realistic collapse leads to a
locally timelike singularity}.

{\item{[4]}}The geometric optics approximation originated in the following works: S. W. Hawking, Comm. 
Math. Phys. {\bf 43} (1975) 199; B. S. DeWitt, Phys. Rep. {\bf 19C} (1975) 295; G.T. Moore,
J. Math. Phys. {\bf 9} (1979) 2679; A review of both techniques may be found in N. D. Birrel and 
P.C.W. Davies, {\it Quantum Fields in Curved Space}, Cambridge Monographs in Math. Phys., Cambridge 
University Press, London (1982).

{\item{[5]}}L. H. Ford and Leonard Parker, Phys. Rev. {\bf D17} (1978) 1485.

{\item{[6]}}W. A. Hiscock, L. G. Williams and D. M. Eardley, Phys. Rev. {\bf D26} (1982) 751.

{\item{[7]}}S. Barve, T.P. Singh, Cenalo Vaz and L. Witten, Nucl. Phys. {\bf B532} (1998) 361;
{\it ibid}, Phys. Rev. {\bf D58} (1998) 104018.

{\item{[8]}}Cenalo Vaz and Louis Witten, Phys. Letts. {\bf B442} (1998) 90.

{\item{[9]}}P. S. Joshi and T. P. Singh, Phys. Rev. {\bf D51} (1995) 6778 and refs. therein;
T. P. Singh and P. S. Joshi, Class. Quant. Grav. {\bf 13} (1996) 559.

{\item{[10]}}R. C. Tolman (1934) Proc. Nat. Acad. Sci. USA {\bf 20} 169; H. Bondi (1947) Mon. 
Not. Astron. Soc. {\bf 107} 410

\bye